\documentclass{ws-ijmpa}
\begin{document}
\markboth{ H. Y\"{u}ksel}{Neutrino Physics from New Measurements}
%\catchline{}{}{}{}{}
\title{NEUTRINO PHYSICS FROM NEW MEASUREMENTS}
\author{\footnotesize  H. Y\"{U}KSEL}
\address{Department of Physics, University of Wisconsin\\
Madison, Wisconsin 53706 USA\\
yuksel@physics.wisc.edu}
\maketitle
\pub{Received (\today) }{}%{Revised (Day Month Year)}
%username=speaker
%password=2d0p0f4
%http://physics.ucr.edu/~billdbrk/Uploads.php?cat=P1&state=ShowSession

\begin{abstract}Emergence of neutrino physics is fueled by the recent growth in
quality and quantity of experimental data, yet, there are still open questions.
How much can the determination of mixing parameters be
improved? Can we improve bounds on $\theta_{13}$ before experiments
designed specifically for this parameter start? How well can we determine or
limit sterile fraction of neutrino flux from sun and sterile mixing
angle? We examine the impact of outcome from new measurements in the context of
providing answers to these questions.
%\keywords{Neutrino Oscillations; Sterile Mixing Angle}
\end{abstract}

\section*{}
It is well established that, solar neutrinos primarily change into another
active flavor\cite{Ahmed:2003kj}. Experimental evidences from
solar\cite{Ahmed:2003kj,Smy:2003jf}, atmospheric\cite{Ashie:2004mr},
reactor\cite{Apollonio:1999ae,Boehm:2001ik,Araki:2004mb} and
long-baseline\cite{Ahn:2002up} experiments can be accommodated into the
framework of neutrino mass, oscillations, and neutrino mixing. In this
formalism, it is known that solar mixing angle,
$\theta_{12}$, is large\cite{Balantekin:2003dc,Balantekin:2003jm}, but not
maximal, and atmospheric mixing angle, $\theta_{23}$, is large, even maximal.
Matter effects probably plays a role in this transformation\cite{Fogli:2003vj}.

\begin{figure}[t]
\begin{minipage}[t]{0.58\linewidth}\vspace*{0.2cm}\hspace*{0.5cm}
\includegraphics[width=0.9\textwidth,height=5cm]{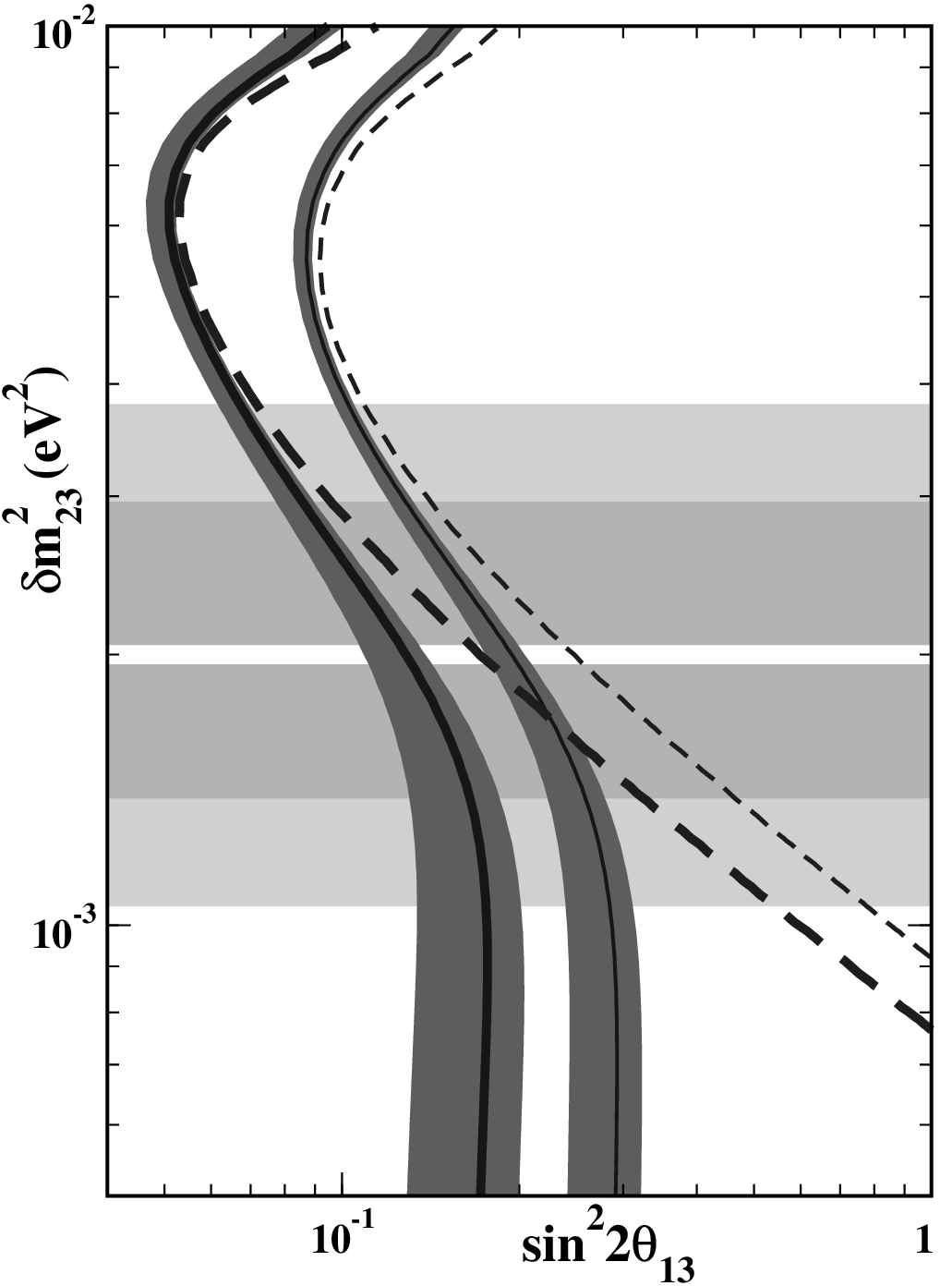}
\caption {
The horizontal shaded regions are 90\%  C.~L. and 99\% C.~L. bounds on
$\delta m^2_{23}$ from SuperK atmospheric data.
The thick (thin) dashed curve is the 90\% C.~L. (3$\sigma$) CHOOZ bound. The
solid curves are bounds from from CHOOZ, KamLAND and solar data at 90\% C.~L.
 and 3$\sigma$ and  the shaded curved bands show the possible impact of
new NC and CC measurements at each confidence level.
} \label{fig:1}
\end{minipage}
\hfill
\begin{minipage}[t]{0.38\linewidth}\vspace*{0.2cm}
\includegraphics[width=1\textwidth,height=5cm]{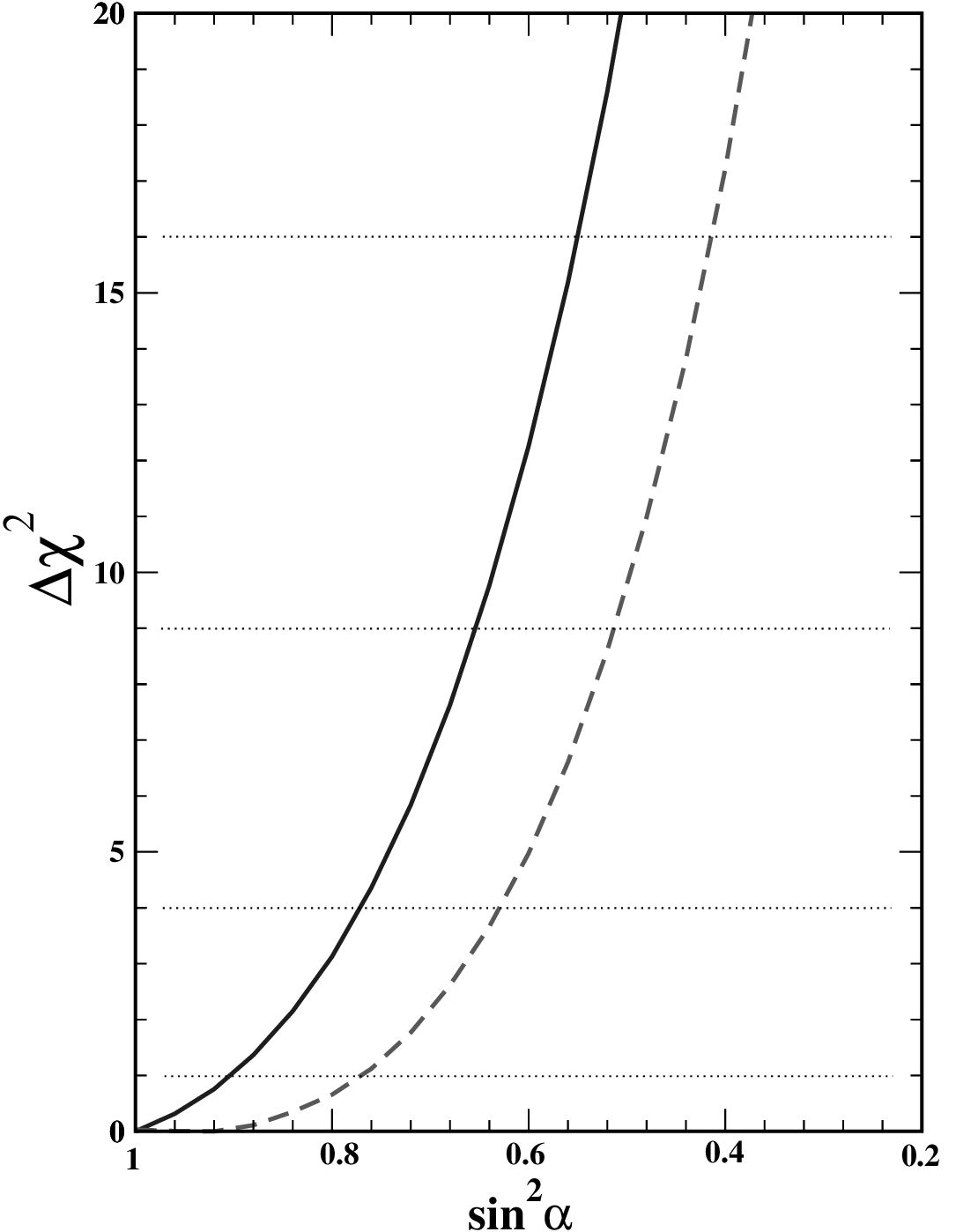}
\caption {$\Delta \chi^2$ vs. $\sin^2 \alpha$ from analyses, all solar and
KamLAND data (solid line) and just SNO and KamLAND data (dashed line). The
horizontal dotted lines are  at the $\Delta \chi^2 =
n^2$ indicating n-$\sigma$ confidence levels. $^8$B flux is free.}
\label{fig:2}
\hfill
\end{minipage}
\end{figure}

The size of the third mixing angle, $\theta_{13}$, is currently best limited by
combined CHOOZ reactor experiment and SK atmospheric data.
In Fig.~\ref{fig:1}, we present bounds on $\theta_{13}$ under certain
considerations. The horizontal shaded regions are 90\%  C.~L. and 99\% C.~L.
bounds on $\delta m^2_{23}$ from SuperK atmospheric data.
The thick (thin) dashed curve is the 90\% C.~L. (3$\sigma$) CHOOZ bound.
The 99\% C.~L. region on $\delta m^2_{23}$ extends as low as
$1\times10^{-3}$eV$^2$. At the lower end of this region, $\theta_{13}$ is
poorly constrained by CHOOZ. The solid curves in Fig.~\ref{fig:1} are bounds
from from CHOOZ, KamLAND and solar data at 90\% C.~L. and 3$\sigma$. Our
analysis shows that, the upper bound on $\theta_{13}$ from KamLAND and solar
neutrino data gets stronger (especially in the region with small $\delta
m^2_{23}$ where
CHOOZ reactor bound is relatively weak), and even dominates as this data gets
refined. In order to examine the possible impact of new NC and CC measurements
from NCD phase of SNO experiment, we add two new hypothetical measurements
with 1-$\sigma$ uncertainties of 5.5\% and 6.4\%, and vary the
corresponding central values by 1-$\sigma$ around their salt-phase
determinations. The shaded curved bands, showing how 90\% C.~L. and 3$\sigma$
bounds vary upon addition of these hypothetical data, suggest that the impact
could be further improvement of the bounds on $\theta_{13}$.

In simple 4-neutrino mixing
scenarios, electron neutrinos change into some linear
combination of active (non-electron) and sterile neutrinos,
\begin{equation}\label{eq:}
\nu_e \rightarrow \cos\alpha\,\nu_s + \sin\alpha\,\nu_{\mu,\tau}\,.
\end{equation}
This is based on the assumption that the electron flavor is distributed among
two eigenstates only. The sterile admixture is described by sterile mixing
angle, $\alpha$. The fraction of oscillating neutrinos changing into active
(sterile) flavor is given by $\sin^2\alpha$ ($\cos^2\alpha$). The average
survival probability of electron neutrinos, $P_{e}$, for SNO experiment can
be calculated as,
\begin{equation}\label{eq:}
P_{e} =
\frac{\int\! \lambda(E_\nu)\sigma_{CC}^{SNO}(E_\nu)
P_{e\rightarrow e}(E_\nu,\delta m^2_{12},\theta_{12} )  dE_\nu}
{\int\! \lambda(E_\nu)\sigma_{CC}^{SNO}(E_\nu) dE_\nu }\,,
\end{equation}
where $\lambda$ is $^8B$ spectrum and $\sigma_{CC}^{SNO}$
is CC cross section. The energy dependent survival probability, $P_{e\rightarrow
e}$, is a function of both neutrino energy, $E_\nu$, and mixing parameters,
$\delta m^2_{12}$ and $\theta_{12}$, which can be taken from KamLAND
experiment. Matter effects has little effect on calculation of $P_{e}$ in the
model we
applied here (especially when sterile fraction is small).
SNO experiment measures CC, $\Phi_{CC}\sim P_{e} \Phi_{^8B}$, and NC,
$\Phi_{NC}\sim P_{a} \Phi_{^8B}$ where $P_{a}$ is active fraction of the
total flux and $\Phi_{^8B}$ is solar ${^8B}$ neutrino flux.
It is estimated that $P_{e}$ can be known up to 7\%
uncertainty\cite{Bahcall:2002zh}, while $\Phi_{NC}$ and $\Phi_{CC}$ can be
measured with 5.5\%, 6.4\% uncertainties in NCD phase of SNO\cite{mc}.
The fraction of oscillating neutrinos changing
into active flavor can be cast into the form,
\begin{equation}
\sin^2 \alpha= {\Phi_{NC}-\Phi_{CC} \over \Phi_{^8B} - \Phi_{CC}}\,.
\end{equation}
We examine the fractional error, $\sigma_{\sin^2 \alpha}/\sin^2 \alpha$, which
requires access to the value of  $^8B$ flux. Adapting SSM value, which has over
20\% uncertainty, yields almost 30\% fractional uncertainty on $\sin^2 \alpha$.
An alternative way is extracting, $^8B$ flux from SNO CC measurement, through
the relation $\Phi_{^8B}=\Phi_{CC}/ P_{e}$. We see that fractional
uncertainty on $\sin^2 \alpha$ can be reduced below 20\% by this
substitution\cite{Balantekin:2004hi}. This result is independent of solar model
uncertainties.

Another method to determine the uncertainty on $\sin^2 \alpha$ is to
perform a statistical analysis based on the oscillation formulae, where $\sin^2
\alpha$ is a fit parameter. We present results of such global
analyses in Fig.~\ref{fig:2} projected on $\sin^2 \alpha$ and marginalized over
all other mixing parameters. The results consistent with non existence of
sterile component in solar neutrino flux. The 1-$\sigma$ bound for the analysis,
in which solar and KamLAND data is considered (the solid line
in Fig.~\ref{fig:2}), yields about 10\% uncertainty on $\sin^2
\alpha$. When only SNO and KamLAND data is considered (the dashed line in
Fig.~\ref{fig:2}), the uncertainty on $\sin^2 \alpha$ increases to
about 20\% (which is comparable to the bound we obtained in simpler
approach we utilized earlier, using $P_{e}$ determination from KamLAND and NC,
CC measurements from SNO). To conclude, oscillation of solar neutrinos
into pure sterile flavor is not allowed but the possibility of a
small sterile component in solar flux cannot be eliminated within the current
precision of model dependencies. In the long run, solar model model independent measurements,
like NC determination by SNO experiment, are the preferred way for
studying the sterile neutrino fraction and/or sterile mixing angle in both
simple approaches (like the one we presented) and full statistical analysis
based on the oscillation formulae.

\section*{Acknowledgements}
I thank A.~B.~Balantekin, V.~Barger, D.~Marfatia and
S.~Pakvasa for collaboration and A.~B.~McDonald for
useful conversations. This work was supported in part by the NSF Grant No.
PHY-0244384 and the Wisconsin Alumni Research Foundation.

\end{document}